\documentclass[12pt,preprint]{emulateapj}

\newcommand{\spitzer}{\emph{Spitzer }}
\newcommand{\microns}{$\mu$m}
\newcommand{\etal}{et al.}

\slugcomment{}

\shorttitle{Primary transit of the planet HD189733b at 3.6 and 5.8 $\mu$m}

\shortauthors{Beaulieu et al.}

\begin{document}

\title{Primary transit of the planet HD189733b at 3.6 and 5.8 $\mu$m}

\author{
J.P. Beaulieu\altaffilmark{1,2},
 S. Carey\altaffilmark{3}, I. Ribas\altaffilmark{4,2}, G. Tinetti\altaffilmark{5,2} }

\altaffiltext{1}{Institut d'astrophysique de Paris, CNRS (UMR 7095), Universit\'e Pierre \& Marie Curie, Paris, France }
\altaffiltext{2}{HOLMES collaboration}
\altaffiltext{3}{IPAC-Spitzer Science Center, California Institute of Technology, Pasadena, California 91125, USA}
\altaffiltext{4}{Institut de Ci\`encies de l'Espai (CSIC-IEEC), Campus UAB, 08193 Bellaterra, Spain}
\altaffiltext{5}{European Space Agency/University College London, Gower Street, London WC1E 6BT, UK}

\begin{abstract}
The hot Jupiter HD 189733b was observed during its primary transit using
the Infrared Array Camera on the \spitzer\ \emph{Space Telescope}. The
transit depths were measured simultaneously at 3.6 and 5.8 $\mu$m. Our
analysis yields values of $2.356 \pm 0.019$\% and $2.436 \pm 0.020$\% at 3.6
and 5.8 $\mu$m respectively, for a uniform source. We estimated
 the contribution of the limb-darkening and star-spot
effects on the final results. We concluded that although the limb darkening increases
by $\sim $ 0.02-0.03 \% the transit depths, and the differential effects
between the two IRAC bands is even smaller, 0.01 \%.
Furthermore, the host star is known to be an active spotted K star with
observed photometric modulation. If we adopt an extreme model of 20 \% coverage
with spots 1000K cooler of the star surface, it will make the observed
transits shallower by 0.19 and 0.18 \%. The difference between the two
bands will be only of 0.01 \%, in the opposite direction to the limb darkening
correction.
If the transit depth is affected by limb darkening and spots,
the differential effects between the  3.6 and 5.8 $\mu$m  bands
are very small.
The differential transit depths at 3.6 and 5.8 $\mu$m
and the recent one published by Knutson et al.( 2007) at 8 \microns\ are in
agreement with the presence of water vapour in the upper atmosphere  of the planet.
This is the companion paper to Tinetti et al. (2007b), where the detailed atmosphere models
are presented.
\end{abstract}

\keywords{Stars: planetary systems --- planetary systems: formation ---{\it Facilities:} \facility{Spitzer}}

\section{Introduction}
\label{sec:intro}

Over 240 planets are now known to orbit stars different from our Sun
(Schneider 2007; Butler et al. 2007). This number is due to increase
exponentially in the near future thanks to space-missions devoted to the
detection of exoplanets and the improved capabilities of the ground-based
telescopes. Among the exoplanets discovered so far, the best known class
of planetary bodies are giant planets (EGPs) orbiting very close-in
(hot-Jupiters). In particular, hot-Jupiters that transit their parent
stars offer a unique opportunity to estimate directly key physical
properties of their atmospheres (Brown, 2001).
In particular, the use of the primary transit (when the planet passes in front of its parent
star) and transmission spectroscopy to probe the upper layers of the
transiting EGPs, has been particularly successful in the UV and visible
ranges (Charbonneau et al. 2002; Richardson et al. 2003, 2003a; Deming et
al. 2005a; Vidal-Madjar et al. 2003, 2004; Knutson et al. 2007; Ballester,
Sing, \& Herbert 2007; Ben-Jaffel, 2007) and in the
thermal IR (Richardson et al. 2006; Knutson et al. 2007).

The hot-Jupiter HD 189733b (Bouchy et al. 2005) has a mass of $M_p=1.15
\pm 0.04$~$M_{\rm Jup}$, a radius of $R_p=1.26 \pm 0.03$~$R_{\rm Jup}$,
and orbits a main sequence K-type star at a distance of 0.0312 AU. This
 exoplanet is orbiting the brightest and closest star discovered so far,
making it one of the prime targets for observations (Bakos et al. 2006;
Winn et al. 2007;  Deming et al. 2006; Grillmair et al. 2007; Knutson et
al. 2007).

Tinetti et al. (2007a) have modeled the infrared transmission spectrum of
the planet HD 189733b during the primary transit and have shown that
\spitzer observations are well suited to probe the atmospheric
composition and, in particular, constrain the abundances of water vapor
and CO. Here we analyze the observations of HD 189733b with the Infrared
Array Camera (IRAC Fazio et al., 2004) on board the \spitzer Space Telescope in two bands
centered at 3.6 and 5.8 $\mu$m. We report the data reduction, and discuss
our results in light of the theoretical predictions and the recent
observation at 8 $\mu$m (Knutson et al. 2007).

\section{Methods}
\label{sec:obs}

\subsection{The observations}

HD 189733 was observed on October 31, 2006 (program id 30590) during the
primary transit of its planet with the IRAC instrument.
During the 4.5 hours of observations, 1.8 hours were spent on the
planetary transit, and 2.7 hours outside the transit. High accuracy in the
relative photometry was obtained so that the transit data in the two bands
could be compared. During the observations, the pointing was held constant to
keep the source centered on a given pixel of the detector. Two reasons
prompted us to adopt this approach:
\begin{itemize}
\item[-] The amount of light detected in channel 1 shows variability that
depends on the relative position of the source with respect to a pixel
center (called the pixel-phase effect). This effect could be up to 4\%
peak-to-peak at 3.6 $\mu$m. Corrective terms have been determined for channel
1 and are reported in the IRAC Manual, but also by Morales-Calder\'on et
al. (2006; hereafter MC06). These systematic effects are known to be
variable across the field. At first order we can correct them using the
prescriptions of MC06 or of the manual, and then check for the need of
higher order corrections.
\item[-] Flat-fielding errors are another important issue. Observations at
different positions on the array will cause a systematic scatter in the
photometric data which may swamp the weak signal we are aiming to detect.
\end{itemize}
Therefore, to achieve high-precision photometry at $3.6~\mu$m, it is
important to keep the source fixed at a particular position in the array.
Staring mode observations can keep a source fixed within 0.15 arcsecond.
It is crucial to have pre-transit and post-transit data to estimate the systematic
effects and to understand how to correct for them.

There is no significant pixel-phase effect for channel 3 of IRAC.
However, the constraint on the flat-fielding error requires that the
source is centered on the same pixel of the detector during the
observations. A latent buildup may affect the response of the detector as
a function of time. To avoid the saturation of the detector for this $K=5.5$
mag target, a short exposure time was used. The observations were split in
consecutive sub-exposures each integrated over 0.4 and 2 seconds for
channel 1 and 3, respectively.

\subsection{Data reduction}

We used the flat-fielded, cosmic ray-corrected and flux-calibrated data
files provided by the \spitzer pipeline. We treated the data of the two
channels separately. We used the BLUE software (Alard 2007) which performs
PSF photometry. Below we describe the details of this approach.

The PSF was reconstructed from a compilation of the brightest, unsaturated
stars in the image. Once the local background had been subtracted and the
flux normalized, we obtained a data set representing the PSF at different
locations on the image. An analytical model was fitted to this dataset by
using an expansion of Gaussian polynomial functions. To fit the PSF
spatial variations, the coefficients of the local expansion are polynomial
functions of the position in the image. Note that the functions used for
the expansion of the PSF are similar to those used for the kernel
expansion in the image subtraction process. A full description of this
analytical scheme is given in Alard (2000).

The position of the centroid was quantified by an iterative process.
Starting from an estimate, based on the position of the local maximum of
the object, we performed a linear fit of the amplitude and the PSF offsets
$(dx,dy)$ to correct the position. The basic functions for this fit were
the PSF and its first two derivatives. Note that in general the
calculation of the PSF derivatives from its analytical model is
numerically sensitive. We recall also that the first moments are exactly
the PSF derivatives in the case of a Gaussian PSF. This procedure
converges quickly: only few iterations are necessary for an accuracy of
less than 1/100 of a pixel.

We performed photometry on all the frames of ch1 and ch3. We tried two
different approaches: we used a Poisson weighting of the PSF fits and then
the weight maps provided by the \spitzer pipeline. These maps contain
for each pixel the propagated errors of the different steps through the
\spitzer pipeline. The results of these two processing runs are almost
indistinguishable. Systematic trends were present in both channels. Here
we discuss each channel separately.

\section{RESULTS}
\label{sec:results}

\subsection{Channel 1}

\begin{figure}
\includegraphics[angle=0,width=9cm]{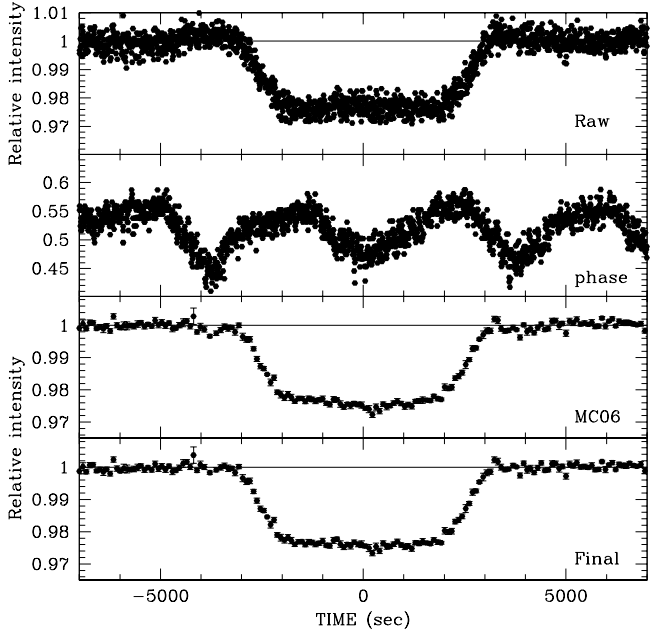}
\caption{Top panel: raw photometric light curve, with systematic trends
due to the pixel-phase effect. Middle panel: variation of the pixel-phase
as a function of time. We can clearly see some correlations between these
two panels, as expected from the known behavior of the IRAC Channel 1.
Bottom panels: data binned by 10 after the correction for the
pixel-phase. In the MC06 labeled panel we corrected the raw photometry
using the prescription of Morales-Calder\'on et al. (2006). In the lowest
panel, we estimated the corrective terms from pre-transit and post-transit
data, and applied them to the full data set.}
\label{fig:fig1}
\end{figure}

Figure \ref{fig:fig1} shows the different steps of the data processing. In
the upper panel we report the raw photometry as produced by BLUE.
Inspection of the lightcurve in the pre-transit and post-transit phases
shows systematic trends with time scales of about 1 hour, with a
peak-to-peak amplitude of 0.7\%, related to the variation of the
pixel-phase due to the jitter of the satellite. The middle panel shows the
pixel-phase. We can clearly see that the flux in the upper panel is correlated
with the pixel-phase. We adopted the MC06 prescription to correct for the
pixel-phase, and show the results binned by 10. Most of the systematic
trends present in the raw photometry are removed, but not entirely. There
are still some trends present for the lowest value of the pixel phase
during the transit, pre-transit, and post-transit. Therefore, we used the
pre and post-transit data to fit the corrective terms, as in the approach
of MC06, and applied them to the full light curve. The results are shown
in the lower panel of Fig. \ref{fig:fig1}. There is an improvement in the
baseline, and also in the transit. The central part of the transit has
still four consecutive
points deviating by 1--2 $\sigma$ around $t \sim 200$~s, corresponding to
the lowest phase value. They are shown on Fig. \ref{fig:fig1} but we will exclude
them from further analysis (Fig. \ref{fig:fig3}).
These remaining systematic effects are due to the phase. We ran calculations both by
including these points and by excluding them, and then compared the
results. Also, we adopted different binning: by 5, 10, 20 or 50 points.
The results are compatible within the errorbars.
As we discuss in more detail below, limb darkening effects
are very small at 3.6 $\mu$m so the shape
of the transit light-curve is box-like.

\subsection{Channel 3}

\begin{figure}
\resizebox{\hsize}{!}{\includegraphics[angle=0,width=9cm]{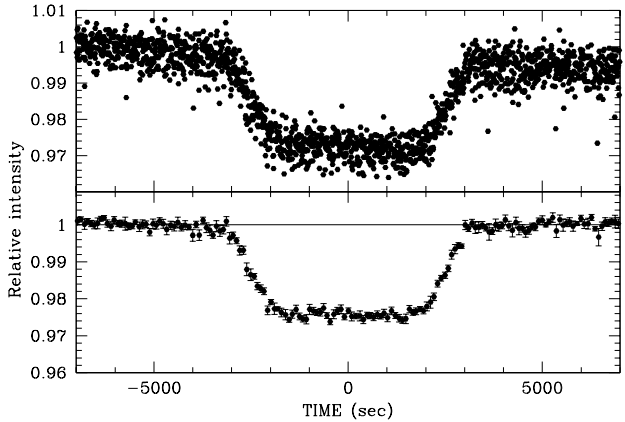}}
\caption{Upper panel: raw photometric light curve with a long term
systematic trend. Lower panel : we estimated the corrective terms from
pre-transit and post-transit data, and applied them to the full data set.
We then bin the data by 10, and estimated the associated errorbar for each
measurement.}
\label{fig:fig2}
\end{figure}

\begin{figure}
\includegraphics[angle=0,width=9cm]{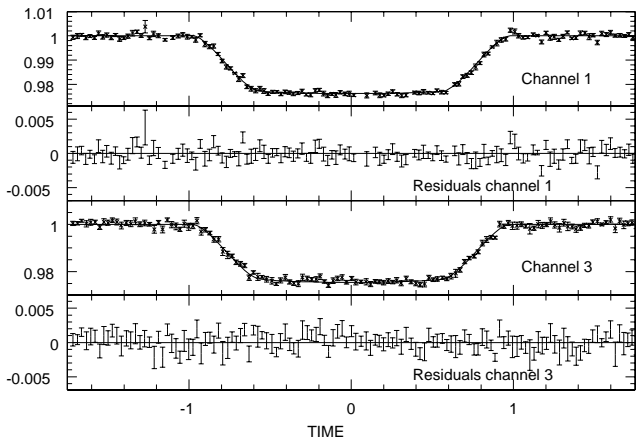}
\caption{From top to bottom: the
channel 1 transit curve with its Mandel and Agol (2002) model, the residuals of the fit,
the channel 3 transit curve with its model, and then the residuals of the fit. }
\label{fig:fig3}
\end{figure}

In Fig. \ref{fig:fig2} we report the raw (upper panel) and the final
(lower panel) photometric data. There is no correlation with the
pixel-phase, but a long term systematic trend can be seen both out of and
in the transit. This trend does not appear to be caused by a latent
buildup, but it is probably linked to the variation of response of the pixels
due to a long period of illumination. We used the pre-transit and the post-transit
data to fit a linear corrective term that we applied to all the data.
The results, binned by 10, are shown in the lower panel of Fig.\ref{fig:fig2}.

\section{DISCUSSION}

\subsection{Comments about the data reduction}

Using the BLUE software we carried out a full modeling of the PSF and obtained
an optimal centroid determination. With an undersampled PSF, in data sets with
strong pixel-phase effects (channels 1 and 2 from \spitzer\ ), accurate
centroid determination is a key issue for achieving high precision
photometry.

In order to tackle the systematic trends that are present in IRAC observations,
it is important to have sufficient baseline observations to analyze
transit data. Here, it has been vital to have sufficient pre-transit and
post-transit data in order to be able to check the nature of the systematics
and correct for them.  The 4.5 hours of observations were centered on 1.8 hours
transit. Given the $\sim 1$ hour time scale of the pixel-phase variations, this
was well adapted, but it is clearly a lower limit on the necessary observing time
scale to achieve such observations.

\subsection{Calculation of the transit depth}

We decide to do a direct comparison between the out-of-transit flux
(Fig. \ref{fig:fig1} and  Fig. \ref{fig:fig2}) and the in-transit flux
(the central 3500 sec) averaged over its flat part for each channel.
We estimate the weighted mean and its error both in the out-of-transit flux
and the in-transit flux.
For channel 1, we have excluded the measurements obtained at the lowest pixel
phase  values as discussed in 3.1.
It yields values of $2.356 \pm 0.019$\% and $2.436 \pm 0.020$\% in
the 3.6- and 5.8-\microns\ bands, respectively (fig.
\ref{fig:absorptions}). This is the same approach as adopted by Knutson et al. (2007).

\subsection{Contribution of limb-darkening}

As a further refinement in our analysis we considered the effects of limb
darkening. By inspection,  the transit is clearly
flat bottomed, and limb darkening was expected to be negligible.
However, it was deemed worth calculating its contribution
because of the high accuracy claimed in the transit
depth measurement given above. We adopted a non-linear limb darkening law model
as described in Mandel \& Agol (2002) to calculate a limb-darkened light
curve. We considered the more sophisticated form using four coefficients
($C_1$, $C_2$, $C_3$, and $C_4$), These were calculated using a Kurucz
(2005) stellar model ($T_{\rm eff}=5000$~K, $\log g=4.5$, solar
abundance), which matches closely the observed parameters of HD\,189733,
convolved with the IRAC passbands. Parameters are given in Table 1.

\begin{deluxetable}{rllll}
    \tablecaption{Limb darkening coefficients \label{tab:LD}}
    \tablehead{
    \colhead{IRAC} & \colhead{ C1 } & \colhead{ C2 } & \colhead{ C3 } & \colhead{ C4 }}
    \startdata
    {3.6~\microns} & 0.6023 & -0.5110 & 0.4655 & -0.1752 \\
    {5.8~\microns} & 0.7137 & -1.0720 & 1.0515 & -0.3825 \\
    \enddata
\end{deluxetable}

A multi-parameter fit of the two light curves using the adopted non-linear
limb-darkening model yielded depths of $2.387 \pm 0.014$\% and $2.456 \pm
0.017$\% in the 3.6- and 5.8-\microns\ bands, respectively. Two small
effects can be identified. First, the limb-darkened transits become some
0.02--0.03\% deeper than those measured assuming a uniform stellar disk.
But also, and very importantly in our case, the relative transit depth
varies by no more than 0.01\%, which is actually about 1/2 of our quoted
error bars. In conclusion, the influence of limb darkening in our
measurements is not significant.

\begin{deluxetable}{lcc}
    \tablecaption{Fitting Parameters of the Transit Curves \label{tab:param}}
    \tablehead{
    \colhead{Parameter} & \colhead{3.6~\microns} & \colhead{5.8~\microns}}
    \startdata
    $R_p / R_\star$(LD) & $ 0.15285 \pm 0.0003$    & $0.1545 \pm 0.0004$ \\
    $b$                 & $0.620 \pm 0.01$      & $0.620 \pm 0.01$   \\
    $(R_p/R_\star)^2$ \% (Uniform)  & $2.356 \pm 0.019$  & $2.436 \pm 0.023$   \\
    $(R_p/R_\star)^2$ \% (LD)       & $2.383 \pm 0.014$  & $2.457 \pm 0.017$  \\
    \enddata
\end{deluxetable}

\subsection{Contribution of spots}

HD 189733 is known to be a relatively active star (Winn et al., 2007, Pont et al. 2007a),
with spots that can cause variations of $\sim$3\% at visible wavelengths
(Strassmeier et al. 2000). These likely arise from rotational modulation over
a period of 12.04 days. To set the context, the observed light variations
would be equivalent to a dark spot with a radius of 1 $R_{\rm Jup}$ and
effective temperature $\sim 1000 K$ cooler than the photospheric effective
temperature of the star. The effect of spots is expected to be
particularly important at visible wavelengths, because the the contrast
with the surrounding photosphere is larger.

In our case the important issue is the possible impact of spots on the
determination of the planetary transit depth. Furthermore, the
effect would be inexistent if spots were distributed homogeneously on the
stellar surface. However, numerous surface maps of active stars, mostly
obtained with the so-called Doppler tomography technique (e.g.,
Strassmeier 2002, and references therein), have revealed that active stars
tend to have an accumulation of active areas (i.e., dark spots) at polar
latitudes. A possible scenario is one in which the planet path during the
transit occurs over an unspotted area of the star and therefore the
resulting transit would appear deeper (since the occulted area would
correspond to the brighter photosphere). Below we analyze this extreme
situation and evaluate the effect on the observed differential depth of
the two channels.

To address this issue we modeled the effects of spots in the bands of
3.6~\microns\ and 5.8~\microns. We took the stellar parameters described
in \S 3.1 and $T_{eff}=3500 K$, $\log g=4.5$ for the spots. We adopted the
NextGen atmosphere models (Hauschildt et al. 1999), based on the PHOENIX
code. The integrated stellar flux was simulated by adding the fluxes from
the photospheric and spotted regions with the appropriate weights to
consider different spot areal coverages. We then calculated photometry by
convolving with the IRAC passbands. Tests using this extreme model and a
20\% surface spot coverage indicate that one could expect an absolute
effect of about 0.19\% in the measured transit depth at 3.6~\microns and
about 0.18\% at 5.8~\microns. Both would be in the sense of making the
spot-corrected transit shallower. As can be seen, while the correction is
large in absolute terms, the difference between the two bands is of about
0.01\% (in the direction of making them more different than measured),
which corresponds to approximately 0.5$\sigma$ of our quoted error bars.
Larger spot coverages would also imply larger effects. For example, a
stellar surface covered 50\% with spots would result in a increased
difference between the two bands by 0.05\%, but this is a very extreme --
and possibly unphysical -- scenario.

It is interesting to evaluate the effect of spot modulation when combining
multi-epoch transit depth measurements. This is relevant in our case
because we also consider the 8~\microns\ measurement by Knutson et al.
(2007), which was obtained with a difference of one orbital period (2.2
days). Using a spot modulation amplitude of $\sim$0.03 mag it can be
deduced that the spot coverage of the stellar hemisphere in view may have
changed by about 2\% during this time lapse. In the case of the 8~\microns\
band the correction for such spot change in the transit depth will be
around 0.01-0.02\% (in either direction depending on whether the spot
coverage has increased or decreased). Again, this is a small correction
(less than 1$\sigma$) that corresponds to an extreme scenario. The effect
is therefore negligible.

Our model also predicts that the effects in the optical wavelengths can be
much larger (of the order of 0.5\% or more in the observed transit depth).
Therefore, the observed difference between the IR and visible radii might
be due to stellar activity. Tinetti et al. (2007a) propose the presence of
optically thick (in the visible) clouds/haze in the upper atmosphere as a
possible explanation of this difference. Additional -- and possibly
simultaneous -- observations in both the visible and IR are needed to draw
firmer conclusions and to disantangle these two potential contributions.

\subsection{Comparison with previous analysis of the same data set}

In an earlier letter, Ehrenreich et al. (2007) adopted a method to analyse the IRAC data that
differs from the standard ones reported in the literature and used here.
In their paper, they conclude that systematic effects contaminate the
final results in such a way that the error bars of Tinetti et al. (2007b)
and Knutson et al. (2007) are severely underestimated. According to the
authors, the accuracy in the radius determination in the IR is not of
sufficient accuracy for the spectroscopic characterization of close-in
atmospheres. In the remainder of this section we argue that it is not the
IR data, and \spitzer in particular, that have insufficient quality but
rather that the lack of accuracy can be attributed to shortcomings in the
analysis method devised by Ehrenreich et al.
As discussed below and as evidenced by the residuals in published Spitzer
photometry  (Knutson et al. 2007, Deming et al., 2007,
Morales-Calder\'on et al., 2006), the systematic effects in the data can be
estimated and robustly corrected for.

Firstly, because of a non-optimal centroid determination, Ehrenreich et
al. found a larger sensitivity to pixel phase effects than the reduction
presented here. Secondly, these authors expressed concern about the
potential contamination by a hot pixel located 4 columns away from the
central pixel of the PSF. With our adopted weighted PSF fit, this pixel
was flagged and thus it is not contaminating the measurement. A third
concern raised by Ehrenreich et al. was the effect of limb darkening on
the measurement of the planet radius. We refer the reader back to \S 4.3
for detailed discussion on this point, but, in short, differential effects
between the 3.6 and 5.8 $\mu$m bands are below 0.01\%, i.e., 0.5 sigma. We
further note that Ehrenreich et al. did not consider the contribution from
star spots while this is of the same order as limb darkening effects, and
especially important when comparing with the results in the visible.

Contrary to the statement of Ehrenreich et al., the 3.6 $\mu$m observations
are not saturated for any part of the observation.  The maximum value of the
peak pixel for the 3.6 $\mu$m data is 115000 electrons which is
well below the published well depth of 145000 electrons.  If the observations
were strongly saturated, we would expect the measured flux density to be
10\% less than the predicted flux density of 1.807 Jy.  Our measured flux at
the beginning of the observation is 1.799 Jy which is in good agreement with
expectations.

We are also at variance with the suggestions of Ehrenreich et al. about
the need to observe multiple transits with the longest possible
out-of-transit baseline. It has not yet been demonstrated that residual
systematics would be reduced by analyzing multiple epochs.  It is very likely
that the pixel-phase effect will have to be treated separately for each
epoch as the acquisitional pointing accuracy of \spitzer is $\sim$ 1
arcsecond (a good fraction of an IRAC array pixel).  The ability to coadd
multiple epochs of IRAC data  will tested with upcoming observations of
HD~209458b as part our \spitzer program WETWORLD (PI Tinetti), or,
alternatively, by using multiple epochs obtained by other groups.
Moreover, it is known that once the satellite settles in a repeatable
jitter pattern, the timescale of systematic effects is of the order of
$\sim$1 hour. Thus, 2.5 hours of pre-transit observations followed by 2
hours of post transit would be our recommended observing strategy to
carefully estimate and correct for systematic effects.

\subsection{Comparison with previous observations}
\begin{figure}
\resizebox{\hsize}{!}{\includegraphics[width=9cm,angle=90]{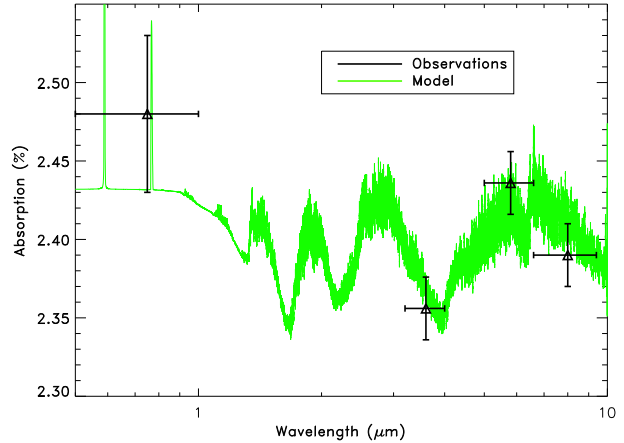}}
\caption{Transit depths as a function of wavelength: our two measurements
at 3.6 and 5.8~\microns\ are indicated with their
error-bars. For comparison we show previous measurements at 8~\microns\
(Knutson et al. 2007)  and in the visible (Winn et al.
2007). Horizontal bars illustrate the instrument
bandpasses. The solid line shows the simulated absorption spectrum of the planet
between 0.5 and 10 $\mu$m. The atmospheric model includes water with a mixing ratio of
$5~10^{-4}$, Sodium and Potassium absorptions, and hazes at the millibar level in the visible.
The underlying continuum is given by $H_2-H_2$ contribution which is sensitive to
the temperature of the atmosphere at pressure higher than the $\sim$ bar level.
Details of the haze free model are given in Tinetti et al., (2007b). Here, hazes are
simulated with a distribution of particles peaked at 0.5 $\mu$m size. In
 this example haze opacity mask the atomic and molecular features at wavelength bluer than 1.2 microns. See also Brown (2001) and Pont et al., (2007b).
 }
\label{fig:absorptions}
\end{figure}

The most recent optical and IR measurements of the radius of HD\,189733b
are plotted in Fig.~\ref{fig:absorptions}  with an underlying model similar
to the one presented by Tinetti et al, 2007b, with  the addition of hazes contributing
in the visible. The $b$ and $R_\star$ values at
3.6 and 5.8~\microns\ are consistent with the visible values (Winn et al.
2007).

Our results are consistent, but not overlapping, with the Knutson et al.
(2007) measurement at 8~\microns. The three primary transit observations
at 3.6, 5.8 and 8~\microns\ with IRAC are in agreement with the
predictions of Tinetti et al. (2007a) and the presence of water vapor in
the atmosphere of the planet. This explanation is not necessarily in
contradiction with the most recent observation of HD 189733b with the
Spitzer Infrared Spectrograph (IRS) using the occultation -- as opposed to the
transit -- (Grillmair et al. 2007). Fortney \& Marley (2007) pointed out that
this observation does not agree with the Knutson et al. (2007) secondary
transit measurement at 8~\microns, hence the IRS observations might have
some problems in the 7.5--10~\microns\ range. Also, the absence of
atmospheric signatures in the thermal emission spectrum of HD\,189733b
might be explained by an isothermal atmosphere (Tinetti et al. 2007b;
Fortney et al. 2006) whereas the primary transit technique allows us to
probe the atmospheric content independently of the temperature gradient.


\section{CONCLUSIONS}
\label{sec:conclusions}

We estimated accurately the radius of the extrasolar planet HD\,189733b
using its primary transit, at 3.6 and 5.8~\microns. The small error bars
are the result of a high signal-to-noise ratio and weak influence of the
limb-darkening effect in the IR. The planetary radius appears $(1.6 \pm
0.5)$\% larger at 5.8~\microns\ than at 3.6~\microns. The
observations match the predictions by Tinetti et al. (2007a).

Detailed interpretation of these results (Tinetti et al. 2007b) combined
with the 8 ~\microns\ observations (Knutson et al. 2007) confirm that
water vapor is the most likely explanation for the observed photometric
signature in the IR. The comparison with the visible is more complex
because of the possibly important role of star spots.

Our observations show that the combination of the primary transit
technique and comparative band photometry at multiple wavelengths is an
excellent tool to probe the atmospheric constituents of transiting
extrasolar planets. Similar studies and observations should be considered
for other targets, especially with the foreseen {\it James Webb Space
Telescope}, which could observe more distant and smaller transiting
planets.


\acknowledgements We thank the staff at the Spitzer Science Center for
their help. We are very grateful to Christophe Alard for having helped us
in the data reduction phases of \spitzer data.  His optimal centroid
determination has been an important contribution to this analysis.
We thank David Kipping for careful reading of the manuscript, and David Sing for
providing the limb darkening coefficients.
 This work is based on observations made with the \emph{Spitzer
Space Telescope}, which is operated by the Jet Propulsion Laboratory,
California Institute of Technology under a contract with NASA. G.~Tinetti acknowledge
the support of the European Space Agency.
IR acknowledges support from the Spanish Ministerio de Educaci\'on y Ciencia via grant AYA2006-15623-C02-02.
 JPB, IR
and GT acknowledge the financial support of the ANR HOLMES.

%


\end{document}